\documentclass[aip,reprint,jap]{revtex4-1}

\usepackage{graphicx}
\usepackage{amsmath}
\usepackage{epstopdf}

\begin{document}


\title{Performance predictions of a focused ion beam from a laser cooled and compressed atomic beam} 



\author{G. ten Haaf}
\affiliation{Department of Applied Physics, Eindhoven University of Technology, P.O. Box 513, 5600 MB Eindhoven, the Netherlands}
\author{S.H.W. Wouters}
\affiliation{Department of Applied Physics, Eindhoven University of Technology, P.O. Box 513, 5600 MB Eindhoven, the Netherlands}
\author{S.B. van der Geer}
\affiliation{Department of Applied Physics, Eindhoven University of Technology, P.O. Box 513, 5600 MB Eindhoven, the Netherlands}
\affiliation{Pulsar Physics, Burghstraat 47, 5614 BC Eindhoven, the Netherlands}
\author{E.J.D. Vredenbregt}
\affiliation{Department of Applied Physics, Eindhoven University of Technology, P.O. Box 513, 5600 MB Eindhoven, the Netherlands}
\author{P.H.A. Mutsaers}
\affiliation{Department of Applied Physics, Eindhoven University of Technology, P.O. Box 513, 5600 MB Eindhoven, the Netherlands}

\date{\today}

\begin{abstract}
Focused ion beams are indispensable tools in the semiconductor industry because of their ability to image and modify structures at the nanometer length scale. Here we report on performance predictions of a new type of focused ion beam based on photo-ionization of a laser cooled and compressed atomic beam. Particle tracing simulations are performed to investigate the effects of disorder-induced heating after ionization in a large electric field. They lead to a constraint on this electric field strength which is used as input for an analytical model which predicts the minimum attainable spot size as a function of amongst others the flux density of the atomic beam, the temperature of this beam and the total current. At low currents ($I<10$ pA) the spot size will be limited by a combination of spherical aberration and brightness, while at higher currents this is a combination of chromatic aberration and brightness. It is expected that a nanometer size spot is possible at a current of 1 pA. The analytical model was verified with particle tracing simulations of a complete focused ion beam setup. A genetic algorithm was used to find the optimum acceleration electric field as a function of the current. At low currents the result agrees well with the analytical model while at higher currents the spot sizes found are even lower due to effects that are not taken into account in the analytical model.
\end{abstract}


\maketitle 

\section{Introduction}

Miniaturization and functional diversification are the driving forces in the semiconductor industry; features on integrated circuits (ICs) become smaller and complexer \cite{itrs}. To be able to produce these features, tools that can image and modify structures at the nanometer length scale, such as a focused ion beam (FIB), are indispensable. Applications of a FIB include physical sputtering (also known as milling), gas assisted deposition \cite{Utke2008} and secondary ion mass spectroscopy (SIMS)\cite{Phaneuf1999}. As the features on ICs become smaller and complexer, the resolution of focused ion beams should increase.

The ion source most often used in commercial FIB instruments is the $\textrm{Ga}^+$ Liquid Metal Ion Source (LMIS) with a reduced brightness of $10^6$ Am$^{-2}$sr$^{-1}$eV$^{-1}$ and an rms energy spread of 2.1 eV \cite{Hagen2008,Bell1988}. It is capable of producing a 1 pA beam that can be focused to a 5-10 nm spot at 30 keV \cite{Raffa}. Due to the relatively large mass of gallium and thus high sputter yield it is currently the preferred source for milling purposes. However, the gas field ionization source (GFIS) has a much higher brightness (estimated at $2\times10^9$ Am$^{-2}$sr$^{-1}$eV$^{-1}$, \cite{Ward2006}) and lower energy spread (less than 1 eV \cite{Ward2006}), which enables sub-nanometer spot sizes. This source has been demonstrated using helium \cite{Ward2006} and neon \cite{Livengood2011}. Due to the low mass of these atomic species the sputter yield is lower than for example gallium. Furthermore, these ions have a larger penetration depth causing subsurface damage to substrates\cite{Tan2010}. Overall this makes it the best choice for imaging, but less suited for nanomachining.

The LMIS and GFIS, both achieve a high brightness by extracting the ions from a small area, thus having a very large current density at the source. Another way to achieve a high brightness is by extracting the ions from a very cold source, which limits the angular spread of the beam. Such a cold source can be created by means of laser cooling and compression of a gas as was proposed by several authors \cite{Freinkman2004,Claessens2005,Hanssen2006}. The so called ultra-cold ion source (UCIS) produced a beam of ionic rubidium with a reduced brightness of $8\times10^4$ A m$^{-2}$sr$^{-1}$eV$^{-1}$ and a rms energy spread of 0.9 eV\cite{Debernardi2012a}. A complete FIB system was built as well, utilizing the similar magneto optical trap ion source (MOTIS). It was able to focus a 0.7 pA beam of lithium to a spot size of 27 nm \cite{Knuffman2011}. The limitations of these sources in terms of brightness and spot size were inherent to the design of the source. Due to the low diffusion rate in a magneto optical trap (MOT), the current density that can be extracted from the ionization volume is limited, which limits the brightness and the extractable current as well\cite{vdGeer2007}.

Here we report on calculations of the expected performance of a focused ion beam based on photo-ionization of a laser cooled and compressed atomic beam of rubidium. By first creating an atomic beam and then laser cooling and compressing it in two dimensions, the fundamental limitations of the UCIS and MOTIS are overcome. The current density is not limited anymore by the diffusion rate, but is determined by the flux density of the atomic beam before ionization. Our calculations aim to give the expected spot size as a function of the beam current and other relevant experimental parameters. The effects of disorder-induced heating and aberrations of a realistic lens system are included.

\section{Source design \label{source design}}

\begin{figure*}
\includegraphics[scale=1]{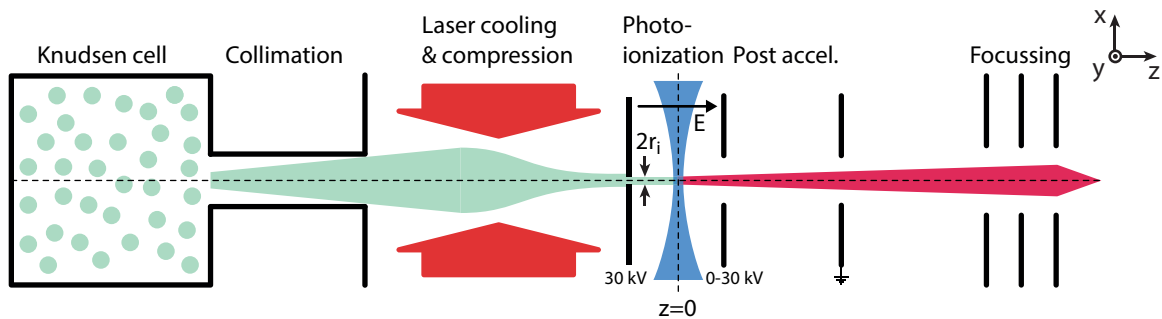}
\caption{Schematic view of the complete beam line of the proposed setup. An atomic beam is created by means of a Knudsen cell with a collimation tube. The beam of rubidium atoms is laser cooled and compressed in the transverse directions. Then it enters the ionization stage through a selection aperture of radius $r_\mathrm{i}$. The selected atoms are photo-ionized and immediately accelerated in an electric field $E$. The ions are further accelerated to an energy of 30 keV in a second acceleration stage and finally focused by a FIB lens system.\label{complete_overview}}
\end{figure*}

The general design of the source under consideration in this article is discussed in \cite{Wouters2014}; here only the most important features are repeated. Figure \ref{complete_overview} shows a schematic overview of the proposed design. An atomic beam of rubidium is created with a Knudsen cell. The advantage of using a Knudsen cell is that it can produce a very high flux of atoms ($>10^{13}$ $\mathrm{s}^{-1}$ at 400 K when using rubidium) as compared to a so called 2D$^+$ MOT \cite{McClelland2013}. In the next stage this beam is laser cooled and compressed in the two transverse directions. Simulations of this magneto-optical compressor (MOC) showed that this increases the flux density $\phi$ of the beam to $4\times10^{19}$ $\mathrm{m}^{-2}\mathrm{s}^{-1}$, while the transverse temperature $T_{\bot}$ of the beam is decreased to 2 mK \cite{Wouters2014}. Note that some results shown here are calculated with a flux density of $5\times10^{19}$ $\mathrm{m}^{-2}\mathrm{s}^{-1}$ and a transverse temperature of 400 $\mathrm{\mu}$K. This lower temperature should be possible when additional sub-Doppler cooling is performed. The effect of different initial flux density and temperature will also be discussed.

Behind the MOC a fraction of the center of the beam is selected with a circular aperture. The transmitted part of the beam is photo-ionized and immediately accelerated in an electric field $E$ in order to suppress disorder-induced heating. The assumptions are made that the complete transmitted beam is ionized and this happens without any increase in transverse velocity spread due to excess energy of the photons. This last assumption is valid since the ionization laser will be tuned near the ionization threshold and most of the excess energy will go to the much lighter electrons. Complete ionization is necessary to convert the high atomic flux density to a high current density. If only a part of the beam would be ionized the current density and thus also the brightness of the beam will be lower. Ionizing the complete beam is possible, although a  build-up cavity will be needed to create a high enough intensity in the ionization laser. Alternatively one can excite the atoms to a Rydberg state and ionize the atoms by means of field ionization\cite{Kime2013}. The current $I$ in the ion beam is set by varying the radius $r_\mathrm{i}$ of the selection aperture, which is given by
\begin{equation}
r_\mathrm{i}=\sqrt{\frac{I}{\pi e\phi}},\label{begin_radius}
\end{equation}
in which $e$ represents the charge of the ions. With the mentioned expected flux density this means the aperture should have a radius of 0.2 $\mathrm{\mu}$m to select a current of 1pA, which is small but not impossible. A problem arising with such a small aperture is the fact that it might get clogged, obstructing the passage of the beam. To prevent this from happening the aperture should be heated in order to evaporate any rubidium which is accumulating on the aperture. After the photo-ionization stage the beam is further accelerated to 30 keV in a second accelerator and finally focused with a FIB lens system.

Since the ionization takes place in an electric field, the position at which an atom is ionized determines its final energy. Therefore the rms energy spread $\sigma_U$ of the beam will be determined by the magnitude of the electric field $E$ and the rms radius $\sigma_\mathrm{L}$ of the ionization laser
\begin{equation}
\sigma_U\approx eE\sigma_\mathrm{L}.\label{energy_spread}
\end{equation}
The approximation made in this equation is that the energy spread of the atomic beam ($\approx$0.04 eV at a Knudsen cell temperature of 400 K) is negligible compared to the energy spread caused by the ionization process. Equation \ref{energy_spread} shows that a smaller electric field leads to a smaller energy spread. Therefore the effect of chromatic aberration of the downstream lens system will be smaller for smaller electric fields. However, the process of disorder induced heating is influenced by the electric field as well since a larger electric field reduces the ion density faster. Thus a large electric field is beneficial to suppress disorder-induced heating but increases the effects of chromatic aberration. This means there exists an optimal electric field which leads to the smallest spot size.

The effects of Coulomb interaction in the UCIS \cite{vdGeer2007} and the MOTIS \cite{Steele2011} were investigated in the past. The current density in an ion beam created from a magneto optical compressed thermal atomic beam is expected to be higher by a factor of $\approx$100 however, making the effects of inter-ion Coulomb interaction even larger. The next section shows the results of particle tracing simulations to investigate the process of disorder-induced heating in the proposed setup and to determine the electric field in the ionization stage which is needed to sufficiently suppress disorder-induced heating for different beam currents. This field determines the energy spread of the beam, which is used in analytical calculations of the minimum achievable spot size shown in section \ref{calculations}. The spot size calculations were verified with particle tracing simulations of a complete and realistic ion beam line shown in section \ref{Genetic_algorithm_optimization}.

\section{Disorder-induced heating\label{heating_simulations}}

The effect of the Coulomb forces between the ions in the beam can be split up into two categories: the space charge effect and statistical effects. Understanding the difference between these two is key to understanding the problems of Coulomb interactions in a focused ion beam. The space charge effect is the effect of the smoothed out average force of all particles. Due to this average effect, the beam will start to expand after it is ionized. The magnitude of this force is correlated to the transverse position of the particle it acts upon. Therefore it can be undone with a positive lens \cite{HoCPO}, which implies that space charge has no effect on the beam's reduced brightness. However, the beam does not consist of a homogeneous space charge, but it contains particles at which the charge is localized. This granularity is the origin of the statistical Coulomb effects in a focused ion beam.

Statistical Coulomb effects can be subdivided into two categories: relaxation of kinetic energy and relaxation of potential energy. Relaxation of kinetic energy occurs when the velocity distribution of the beam is anisotropic, i.e., when the temperature in one direction is different from the temperature in other directions. When this happens the energy present in the random motion in one direction can be transferred to the other directions due to Coulomb collisions. An example of such a process is the well known Boersch effect \cite{boersch}.

When ions are created from a laser intensified atomic beam, their transverse temperature will be of the order of 2 mK. At that moment the longitudinal temperature will be of the order of the temperature of the Knudsen cell. Therefore a process which can be described as the opposite of the Boersch effect can occur, i.e., a relaxation of kinetic energy from the longitudinal to the transverse direction. However, ions are usually accelerated to 30 keV in FIBs, which decreases the longitudinal temperature of the beam\cite{Zimmermann} with a factor $10^6$. Therefore the effects of this process are expected to be minor.

Relaxation of potential energy is also known as disorder-induced heating. When a laser cooled and compressed atomic beam is ionized, ions are created at random initial positions. Therefore the Coulomb interaction forces between these ions will point in random directions and have random magnitudes. In other words, a certain amount of potential energy is created which will relax into random kinetic energy, i.e., the beam heats up. 

Disorder-induced heating has been investigated in the context of ultra-cold plasmas \cite{Killian}. In such systems, thermalization will lead to kinetic energies $k_\mathrm{B}T_\mathrm{f}$ of the order of the initial potential energy
\begin{equation}
k_\mathrm{B}T_\mathrm{f}\approx\frac{e^2}{4\pi\epsilon_0a},\label{equilibration}
\end{equation}
in which $\epsilon_0$ is the vacuum permittivity and $a$ is the Wigner Seitz radius which for a beam is given by
\begin{equation}
a=\left(\frac{3v}{4\pi\phi}\right)^\frac{1}{3},
\end{equation}
in which $v$ is longitudinal velocity of the atoms. The final temperature $T_\mathrm{f}$ is reached on a time scale of the order of the inverse plasma frequency $\omega_\mathrm{p}^{-1}$, which for a beam is given by
\begin{equation}
\omega_\mathrm{p}^{-1}=\sqrt{\frac{mv\epsilon_0}{\phi e^2}},\label{plasmaf}
\end{equation}
in which $m$ represents the mass of the particles. Using these equations for ions created from a laser cooled and compressed atomic beam with a typical flux density of $5\times10^{19}$ m$^{-2}$s$^{-1}$ and atoms traveling at $\approx300$ $\mathrm{ms^{-1}}$, the beam will heat up to $\approx15$ K in about 18 ns. Recalling that the transverse temperature of the atomic beam is expected to be 2 mK, this means the temperature increases with approximately four orders of magnitude, meaning the reduced brightness will decrease with four orders of magnitude.

These numbers clearly indicate the problem of disorder-induced heating. In order to better understand the problem, particle tracing simulations of the ion beam have been performed. Moreover, these particle tracing simulations are used to investigate the effect of experimental quantities such as the electric field $E$ and flux density $\phi$ on the heating process.

\subsection{Simulation Setup}

The process of disorder-induced heating in an accelerating ion beam is investigated with particle tracing simulations using the General Particle Tracer code\cite{gpt}. This code solves the three-dimensional equations of motion for a specified set of particles, in our case individual rubidium ions. It includes externally applied electric fields in the calculation as well as all pairwise Coulomb interactions. Therefore it takes into account all granularity effects as long as the number of particles is chosen sufficiently large, to mimic a continuous beam.

The ions in the simulation are created at random initial positions with random initial velocities, but taking into account certain distributions. The transverse position distribution was taken constant and non-zero for $r\leq r_\mathrm{i}$ and zero for $r>r_\mathrm{i}$, in which $r$ is the radial position of the particle and $r_\mathrm{i}$ is given by equation \ref{begin_radius}. Since the longitudinal initial distribution of the ions will be determined by the ionization laser it was chosen Gaussian with an rms radius $\sigma_\mathrm{L}$ of 3 $\mathrm{\mu}$m and centered at longitudinal position $z=0$. The velocity distributions in the x- and y-direction are Gaussian with a standard deviation $\sigma_{v_\bot}$ given by
\begin{equation}
\sigma_{v_\bot}=\sqrt{\frac{k_\mathrm{B} T_\bot}{m}},
\end{equation}
in which $T_\bot=400$ $\mathrm{\mu K}$ is the transverse temperature achieved with laser cooling and compression. The longitudinal velocity distribution was also assumed to be Gaussian, but with a standard deviation determined by the longitudinal temperature $T_{//}=400$ K and an average $\langle v_z\rangle$ of
\begin{equation}
\langle v_z\rangle=\sqrt{\frac{8 k_\mathrm{B} T_{//}}{\pi m}}.
\end{equation}

The only implemented external electric field was a constant electric field $E$ in the longitudinal direction to accelerate the ions. The results shown in the next subsection are obtained at $z = 10$ mm.

\subsection{Results\label{simulation_results}}

\begin{figure}
\includegraphics[scale=1]{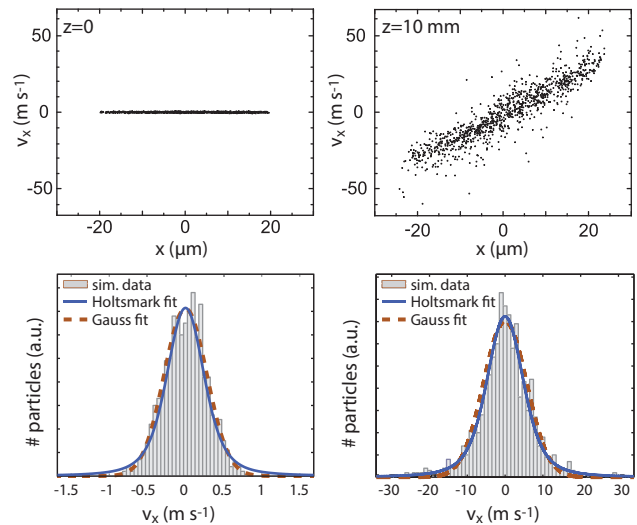}
\caption{Phase space and transverse velocity distribution plots at different longitudinal positions. Both of the velocity distributions are fitted with a Gaussian distribution as well as with a Holtsmark distribution. The distribution at z=0 resembles a Gaussian distribution more than a Holtsmark distribution with a reduced chi squared value of 0.63 for the Gaussian as compared to 1.2 for the Holtsmark distribution. The distribution at z=10 mm resembles a Holtsmark distribution more than a Gaussian distribution with a reduced chi squared value of 0.90 for the Holtsmark distribution as compared to 1.7 for the Gaussian.\label{phasespace}}
\end{figure}

Figure \ref{phasespace} shows a typical simulation result. It shows phase space plots and velocity distributions of a simulated beam at $z$=0 and $z$=10 mm. The simulation was performed with a beam current of 10 nA, a flux density of $5\times 10^{19}$ $\textrm{m}^{-2}\textrm{s}^{-1}$ and an electric field of 1 MVm$^{-1}$. There are two obvious differences between the two phase space plots. First of all, a correlation has developed between the transverse velocity and transverse position. This is caused by the correlation between the space charge force and the positions, which was explained in the beginning of this section. The other difference is the fact that the phase space density, and thus also the brightness, is lower at $z=10$ mm. This clearly visualizes the effect of disorder induced heating.

The velocity distributions shown in figure \ref{phasespace} are obtained from the phase space distributions by fitting a linear function through the phase space data and subtracting this linear function from the data. From this corrected data the velocity distribution is shown. As can be seen, the distribution at $z=0$ is a Gaussian, which was the velocity distribution that was used as input for the simulation. The distribution at $z=10$ mm is much broader than the distribution at $z=0$, which is the effect of disorder-induced heating. Furthermore it resembles a Holtsmark distribution\cite{HoCPO} more than a Gaussian distribution, which can be concluded from the lower reduced chi squared value of the Holtsmark fit as compared to the Gaussian fit. Characteristic for the Holtsmark distribution are the much broader side wings than the Gaussian distribution. These side wings are the reason that the second moment of the distribution does not exist, i.e., the rms radius of the distribution is infinite. Therefore the reduced brightness of a beam with such a velocity distribution will approach zero. However, this does not mean that the peak reduced brightness will approach zero. A better measure for the beam quality is therefore the reduced brightness of 50 percent of the beam $B_\mathrm{r,50}$, which was used in this research. The method to calculate this brightness was shown earlier by Van der Geer et al. \cite{vdGeer2007}.

\begin{figure*}
\includegraphics[scale=1]{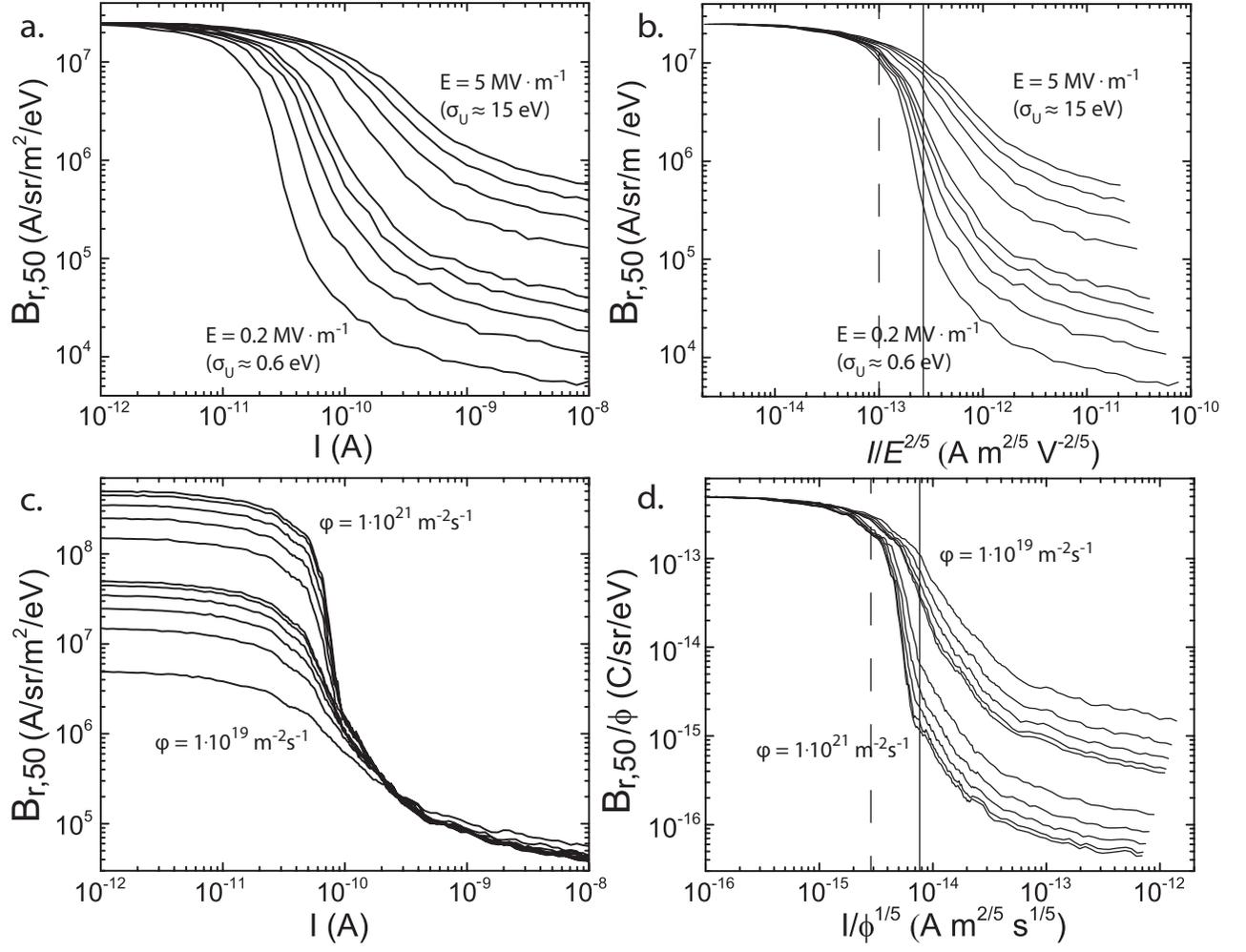}
\caption{Simulation results showing the reduced brightness (a,b,c) of 50 percent of the beam or this value divided by the flux density (d) at a longitudinal position of 10 mm for: (a) a constant flux density of $5\times 10^{19}$ $\textrm{m}^{-2}\textrm{s}^{-1}$ and varying electric field strengths (0.2,0.4,0.6,0.8,1,2,3,4 and 5 MVm$^{-1}$) as a function of the current, (b) a constant flux density of $5\times 10^{19}$ $\textrm{m}^{-2}\textrm{s}^{-1}$ and varying electric field strengths as a function of a scaled current, (c) a constant electric field strength of 1 $\textrm{MV}\textrm{m}^{-1}$ and varying flux densities ($1,3,5,7,9\times10^{19}$ $\textrm{m}^{-2}\textrm{s}^{-1}$, $1,3,5,7,9\times10^{20}$ $\textrm{m}^{-2}\textrm{s}^{-1}$ and $1\times10^{21}$ $\textrm{m}^{-2}\textrm{s}^{-1}$) as a function of the current and (d) a constant electric field of 1 $\textrm{MV}\textrm{m}^{-1}$ and varying flux densities as a function of a scaled current. The solid vertical lines in figure b and d show the current for which the transverse size of the beam is equal to the average initial longitudinal separation of the ions in the beam (equation \ref{scaling}). The dashed vertical lines indicate the end of the pencil beam regime (equation \ref{Iend}).\label{scans}}
\end{figure*}

Simulations were performed with different values for the current, electric field and flux density. The results are shown in figure \ref{scans}. Figure \ref{scans}a shows $B_\mathrm{r,50}$ as a function of the current for electric field strengths ranging from 0.2 MV$\mathrm{m}^{-1}$ to 5 MV$\mathrm{m}^{-1}$. These simulations were performed at a flux density of $5\times 10^{19}$ $\mathrm{m}^{-2}\mathrm{s}^{-1}$. The results for all electric field strengths show similar behavior. At low currents, the brightness stays constant with increasing current. In this current region the beam is in the so-called pencil beam regime, which is characterized by the fact that the transverse size $d=2r_\mathrm{i}$ of the beam is smaller than the average longitudinal separation between individual ions in the beam. In this regime all ions are more or less behind each other instead of next to each other, so that the interaction forces will predominantly point in the longitudinal direction. Therefore, transverse heating will be limited and the brightness unaffected. However, at a certain current, the transverse size of the beam will become too large. At that current heating will also occur in the transverse direction and the brightness will drop. Finally, when the current becomes even larger, a growth of the transverse size has almost no effect on the brightness anymore.

In order to find the electric field needed to suppress disorder-induced heating, the current at which the pencil beam regime ends needs to be identified as a function of the applied acceleration field. Since this current is difficult to quantify the current $I_{\Delta z=2r_\mathrm{i}}$ is calculated at which the transverse size of the beam is equal to the initial average longitudinal separation between the ions. This longitudinal separation $\Delta z$ between subsequent ions is given by
\begin{equation}
\Delta z=\frac{eE}{2m}\Delta t^2+v_\mathrm{a}\Delta t\approx\frac{eE}{2m}\Delta t^2\label{deltaz},
\end{equation}
in which $v_\mathrm{a}$ is the average atomic velocity before ionization and $\Delta t=e/I$ is the average time between subsequent ionizations. Obviously, the approximation made in this equation is only valid if $E\gg\frac{2mIv_\mathrm{a}}{e^2}$. Equating the value in equation \ref{deltaz} to $2r_\mathrm{i}$ and using equation \ref{begin_radius} to solve for the current leads to
\begin{equation}
I_{\Delta z=2r_\mathrm{i}}\approx\left(\frac{e^3E}{4m}\sqrt{\pi e \phi}\right)^\frac{2}{5}.\label{scaling}
\end{equation}

In order to verify this scaling, figure \ref{scans}b shows the same brightness data as figure \ref{scans}a, but now as a function of the scaled current $I/E^{2/5}$. The value shown in equation \ref{scaling} is indicated with a vertical solid line. It can be seen that at this value the brightness is already lower as compared to its initial value for all electric field strengths. However, in all simulations the brightness starts decreasing at approximately the same value of the scaled current.

Figure \ref{scans}c shows $B_\mathrm{r,50}$ as a function of the current for flux densities ranging from $1\times 10^{19}$ $\textrm{m}^{-2}\textrm{s}^{-1}$ to $1\times 10^{21}$ $\textrm{m}^{-2}\textrm{s}^{-1}$. A larger flux density translates to a higher brightness at low currents, i.e., in the pencil beam regime where no transverse heating takes place. However, for large currents, all flux densities approximately lead to the same $B_\mathrm{r,50}$. The loss in brightness due to disorder induced heating is larger for larger flux densities since the particles are initially closer together, i.e., they are created with a larger potential energy. However, at high currents this larger loss in brightness due to heating is compensated by the higher initial brightness. This leads to the observation that the brightness is nearly independent of the initial flux density at large currents. Figure \ref{scans}d shows that the scaling of equation \ref{scaling} also applies to the flux density. However, due to the lower power of $\phi$ in equation \ref{scaling} the effect is less prominent.

The simulation data shown in figure \ref{scans} shows that the current $I_\mathrm{end}$ at which the pencil beam regime ends, scales according to equation \ref{scaling}. However, the exact value of $I_\mathrm{end}$ is difficult to define since there is no hard limit for what is a pencil beam and what is not. Here the limit is obtained empirically from figure \ref{scans}b and is set at
\begin{equation}
I_\mathrm{end}=\frac{E^{2/5}}{\alpha}\left(\frac{\phi}{\phi_0}\right)^{1/5},\label{Iend}
\end{equation}
in which $\alpha=10^{13} \mathrm{A^{-1}V^{2/5}m^{-2/5}}$ and $\phi_0=5\times 10^{19}$ $\mathrm{m}^{-2}\mathrm{s}^{-1}$ is the flux density at which the simulations in figure \ref{scans}b  were performed. The data shows that at this current the brightness has approximately decreased a factor of two. When equation \ref{Iend} is rewritten in terms of the electric field, an equation is obtained for the electric field that is needed to reasonably suppress disorder induced heating as a function of the current
\begin{equation}
E(I)=\left(\alpha I\right)^{\frac{5}{2}}\left(\frac{\phi_0}{\phi}\right)^{\frac{1}{2}}.\label{E(I)}
\end{equation}

\section{Spot size calculations\label{calculations}}

Now the electric field we need to apply to maintain a high brightness is known, calculations can be performed to find the optimal probe size that can be reached with a focused ion beam based on laser cooling and compression. First we will show analytical calculations in which the three most important contributions to the probe size are taken into account: the finite brightness of the beam and spherical and chromatic aberration of a realistic FIB lens system. As a measure for the size of the distribution the diameter $d_{50}$ that contains 50 percent of the current is used. The probe sizes of the three individual contributions and the procedure to add them together will be introduced. Then the total spot size will be optimized for the cases of only chromatic aberration and only spherical aberration by changing the size of the beam at the position of the final lens. Since we use the electric field of equation \ref{E(I)} in the analytical calculations, the results will not include any trade off between a high brightness due to a high electric field and a low energy spread due to a low electric field. However, it is possible that this leads to smaller spot sizes, especially at high currents, where a very high electric field is needed to suppress disorder-induced heating. To investigate this effect particle tracing simulations of the complete ion beam line are performed, combined with a genetic algorithm optimization to find the optimum electric field. These simulations are the subject of section \ref{Genetic_algorithm_optimization}.

The emittance, or brightness, limited spot size $d_{50,B}$ is derived from the definition of the emittance $\epsilon_{i}$, given by
\begin{equation}
\epsilon_{i}=\sqrt{\left\langle i^{2}\right\rangle \left\langle v_{i}^{2}\right\rangle -\left\langle iv_{i}\right\rangle ^{2}},\label{emittance}
\end{equation}
in which $i$ denotes a transverse direction, $v_i$ denotes the velocity in that transverse direction and $\langle.\rangle$ denotes the average over all particles. The assumption is made that no heating of the beam takes place so that emittance is a conserved quantity. This assumption is reasonably justified as long as the applied electric field in the acceleration stage is equal or larger than the value given by equation \ref{E(I)}. In the final waist of the beam after the last lens and in the first waist of the beam at the position of ionization ($z$=0 in figure \ref{complete_overview}) there is no correlation between position and velocity. Therefore the correlation term in equation \ref{emittance} becomes zero at these longitudinal positions. Conservation of emittance and the fact that for an azimuthally symmetric beam $\sqrt{\langle x^2 \rangle}=\sqrt{\langle y^2 \rangle}=\sqrt{\langle r^2 \rangle}/\sqrt{2}$ gives
\begin{equation}
\sigma_{r,\rm{i}}\sigma_{v_r,\rm{i}}=\sigma_{r,\rm{f}}\sigma_{v_r,\rm{f}},
\end{equation}
in which $\sigma_{r,\rm{i}}$ and $\sigma_{r,\rm{f}}$ are the rms radial sizes of the beam in the ionization plane and the focus of the beam and $\sigma_{v_r,\rm{i}}$ and $\sigma_{v_r,\rm{f}}$ are the rms spreads in radial velocity in the ionization plane and the focus of the beam. The current distribution in the ionization plane is assumed to be uniform with a radial size $r_\mathrm{i}$. For such a distribution $\sigma_{r,\rm{i}}=r_\mathrm{i}/\sqrt{2}$. Furthermore, the current density distribution at the final focusing lens is also assumed to be uniform with a radial size $r_\mathrm{L}$. This assumption is justified due to the small transverse velocity spread of the ions. On the basis of geometrical arguments, the rms spread in radial velocity after the final lens with focal length $f$ is now given by $\sigma_{v_r,\rm{f}}=v_zr_\mathrm{L}/(\sqrt{2}f)$. Finally, the rms spread in radial velocity in the ionization plane is given by $\sigma_{v_r,\rm{i}}=\sqrt{2k_\mathrm{B}T_\bot/m}$. Together, these equations lead to
\begin{equation}
\sigma_{r,\rm{f}}=\frac{d_{50,B}}{2}=\frac{fr_\mathrm{i}}{r_\mathrm{L}v_z}\sqrt{\frac{2k_BT_\bot}{m}}.\label{d50Br}
\end{equation}
The first equality in equation \ref{d50Br} is generally valid for a uniform circular distribution.

The probe size contributions of spherical and chromatic aberration are obtained from Wang et al. \cite{Wang1991} and given by 
\begin{equation}
\begin{split}
&d_\mathrm{50,S}=\frac{1}{4\sqrt{2}}C_\mathrm{S}\frac{r_\mathrm{L}^{3}}{f^{3}}\\
&d_\mathrm{50,C}=0.811C_\mathrm{C}\frac{\sigma_{U}}{U_{0}}\frac{r_\mathrm{L}}{f},\label{d50SC}
\end{split}
\end{equation}
in which $U_0$ is the average kinetic energy of the ions in the beam and $\sigma_{U}$ is the energy spread. This energy spread is calculated by inserting the electric field given by equation \ref{E(I)} into equation \ref{energy_spread}. To get realistic results, aberration constants of $C_\mathrm{C}$ = 100 mm and $C_\mathrm{S}$ = 850 mm were used, which are typical for commercial FIB columns having sample tilt capability. Improved performance could be obtained by optimizing the aberration constants for a cold ion source and specified working distance.

\begin{figure}
\includegraphics[scale=1]{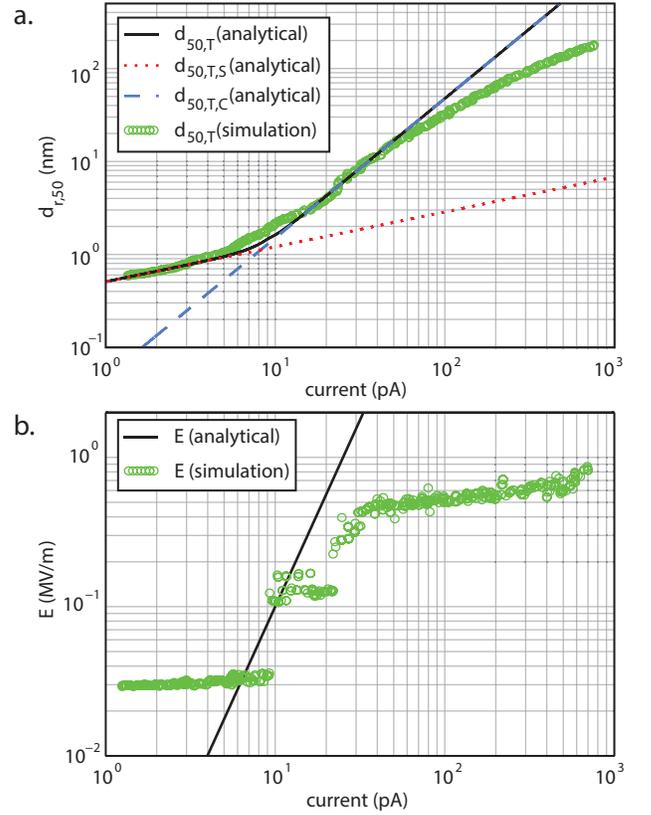}
\caption{(a) Result of the analytical calculation of the spot size as a function of the current (solid line) as well as the result of the genetic optimization of the spot size with particle tracing simulations (circles, the simulations are discussed in section \ref{Genetic_algorithm_optimization}), both for $U_0=$30 keV, $\phi=5\times10^{19}$ $\mathrm{m}^{-2}\mathrm{s}^{-1}$ and $T_{\bot}400$ $\mathrm{\mu}$K. The figure also shows the spherical (dotted line) and chromatic (dashed line) aberration limited spot size as a function of the current. Note that since the flux density is fixed here, the radius of the beam defining aperture is varied when the current is varied due to the relation given by equation \ref{begin_radius}. (b) The electric field used to accelerate the ions in the analytical calculation (solid line) and in the genetic optimization using particle tracing simulations (circles, the simulations are discussed in section \ref{Genetic_algorithm_optimization}) as a function of the current.\label{d50vsI}}
\end{figure}

The total probe size $d_\mathrm{50,T}$ is calculated with the root power sum algorithm of Barth and Kruit\cite{Barth1996}
\begin{equation}
d_\mathrm{50,T}=\left(\left(d_\mathrm{50,S}^{1.3}+d_{50,B}^{1.3}\right)^{\frac{2}{1.3}}+d_\mathrm{50,C}^{2}\right)^{\frac{1}{2}}.\label{d50T}
\end{equation}
An important parameter in minimizing this total probe size is the so called aperture angle $\theta=\frac{r_\mathrm{L}}{f}$. As can be seen in equations \ref{d50Br} and \ref{d50SC} the brightness contribution to the probe size is inversely proportional with $\theta$ while the chromatic aberration contribution is proportional to $\theta$ and the spherical aberration contribution is proportional to $\theta^3$. An analytical optimization of $d_\mathrm{50,T}$ in terms of $\theta$ is complicated, due to the various powers in equation \ref{d50T}. Therefore separate optimizations of the spherical and chromatic aberration limited probe size were performed in which the term containing $d_\mathrm{50,C}$ or the term containing $d_\mathrm{50,S}$ were left out respectively. The results of these optimizations are the spherical aberration limited spot size $d_\mathrm{50,T,S}$ and chromatic aberration limited spot size $d_\mathrm{50,T,C}$ given by
\begin{multline}
d_\mathrm{50,T,S}=\left(3^{\textrm{-}\frac{3}{5.2}}\textrm{+}3^{\frac{1}{5.2}}\right)^\frac{1}{1.3}\left(\frac{2k_\mathrm{B}^3}{\pi^3e^3}\right)^\frac{1}{8}\times C_\mathrm{S}^\frac{1}{4}\left(\frac{IT_\bot}{\phi U_0}\right)^\frac{3}{8}\label{d50TS}
\end{multline}
and
\begin{multline}
d_\mathrm{50,T,C}=\left(\frac{16\times0.811^2k_\mathrm{B}e}{\pi}\right)^\frac{1}{4}\left(\frac{T_\bot\phi_0\alpha^5I^6C_\mathrm{C}^2\sigma_\mathrm{L}^2}{U_0^3\phi^2}\right)^\frac{1}{4}. \label{d50TC}
\end{multline}
Both of these results are shown in figure \ref{d50vsI}a for $U_0=$30 keV, $\phi=5\times10^{19}$ $\mathrm{m}^{-2}\mathrm{s}^{-1}$ and $T_{\bot}=400$ $\mathrm{\mu}$K. It also shows a plot of the minimized complete spot size, given by equation \ref{d50T}. For this plot, either the spherical aberration optimal aperture angle or the chromatic aberration optimal aperture angle was used, depending on which lead to the smallest spot size. Figure \ref{d50vsI}b shows a plot of the electric field used in the calculation.

As can be seen, the probe size will be dominated by spherical aberration below 10 pA, while at higher currents, meaning larger beam radii due to the relation in equation \ref{begin_radius}, chromatic aberration is more important. This behavior is exactly opposite as compared to other sources such as the liquid metal ion source in which the energy spread is independent of the current. For such sources the chromatic aberration limited spot size\cite{Kruit1995} is proportional to $I^\frac{1}{4}$, which is a weaker dependence than the $I^\frac{3}{8}$ dependence of the spherical aberration limited spot size. However, in a focused ion beam based on laser cooling a higher electric field is needed at higher currents to suppress disorder-induced heating as explained in the previous section. Therefore the chromatic aberration limited spot size is proportional to $I^\frac{3}{2}$ as shown in equation \ref{d50TC}. This is the reason why chromatic aberration is dominant at high currents instead of low currents.

The spot size plot shown in figure \ref{d50vsI}a is based on assumptions of the attainable flux density and temperature after laser cooling and compression as explained in section \ref{source design}.	With the analytical model shown in this section it is also possible to test what spot size can be reached with different initial conditions. Figure \ref{d50_vs_phi,T} shows the dependence of the spot size on the initial flux density and temperature of the beam, assuming a 30 keV beam containing 1 pA. Note that since the current is fixed here, varying the flux density of the beam also means varying the beam radius. At a current of 1 pA the spot size is almost completely determined by spherical aberration and brightness so that equation \ref{d50TS} is valid. This equation and figure \ref{d50_vs_phi,T} show that a factor of 10 increase in temperature or decrease in flux density leads to a factor of $10^\frac{3}{8}$ increase in spot size. For higher currents, at which the spot size will be limited by chromatic aberration, a factor of ten increase in temperature or decrease in flux density will lead to a factor of $10^\frac{1}{4}$ or $10^\frac{1}{2}$ increase in spot size respectively. Additional work \cite{Wouters2014} showed that by using a thermal source of rubidium and a compact laser cooling stage a flux density of $\phi=4\times10^{19}$ $\mathrm{m}^{-2}\mathrm{s}^{-1}$ and transverse temperature of $T_{\bot}=2$ mK could be reached. As shown in figure \ref{d50_vs_phi,T}, a 1 pA beam with these characteristics can be focused to 1 nm.
\begin{figure}
\includegraphics[scale=1]{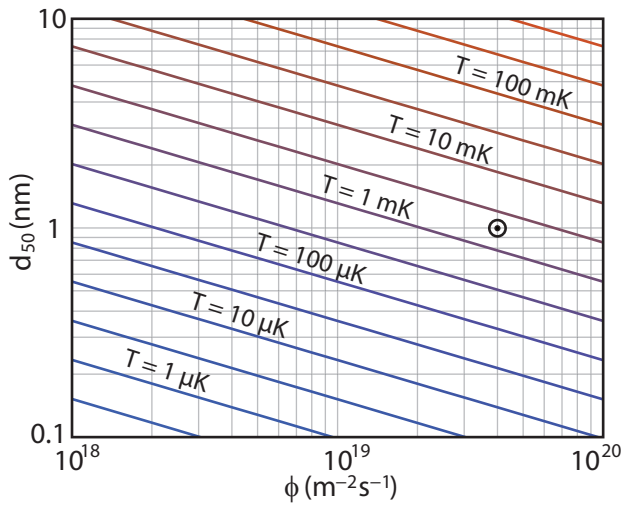}
\caption{This figure shows the dependence of the attainable spot size on the flux density and temperature of the beam after laser cooling and compression, with a 30 keV beam with a current of 1 pA. The dot marks the location of the temperature and flux density we expect to realize in our experimental setup.\label{d50_vs_phi,T}}
\end{figure}

In the analytical calculations shown here the electric field is chosen such that all disorder-induced heating is suppressed within approximately a factor of two decrease in brightness, see section \ref{simulation_results}. Therefore the electric field has a 5/2 power dependence on the current. This is correct for low currents, but from a certain current onward it may not be beneficial anymore to use the electric field according this dependence. It might be better to use a smaller electric field, so that the energy spread is lower, while accepting that some disorder-induced heating takes place. In the optimal situation the electric field should be chosen such that the loss of brightness balances the decrease of energy spread. Since we can not describe the shape of the plots in figure \ref{scans} analytically this effect is investigated numerically.

\section{Genetic algorithm optimization\label{Genetic_algorithm_optimization}}

To investigate the effect described in the last paragraph of the previous section, particle tracing simulations of the complete ion beam line are performed in combination with a genetic algorithm to find the optimal experimental parameters that lead to the smallest spot size. The genetic algorithm is an optimization method based on evolution\cite{Deb}. In each step a certain population of solutions is created. The best of these solutions are selected as parents, which are used to create the input parameters for their children in the next generation. In this way the different solutions evolve towards the optimal solution of the problem. In the case of optimizing the parameters for a focused ion beam there are two main objectives: an as large as possible current in an as small as possible spot size. Therefore a multi-objective genetic algorithm is used. With multiple objectives, the set of best solutions is defined as the set of solutions for which there is no other solution which scores better on one of the objectives without worsening one of the other objectives. This set of solutions is called the pareto front. The optimization shown here is performed using the built-in multi-objective genetic algorithm of Matlab.

\subsection{Simulation Setup}

The simulations in this section are also performed with the GPT code. This time the complete ion beam line is incorporated, including a more realistic accelerator structure and lens column. As explained in section \ref{calculations}, the aperture angle of the last lens must be variable in order to optimize the spot size. This can either be done by varying the length of the system or by changing the divergence of the ion beam before the lens column. In practical sense the last option is preferred, which is incorporated in the simulation with a decelerating einzel lens that focuses the ions after which the beam diverges.

\begin{figure}
\includegraphics[scale=1]{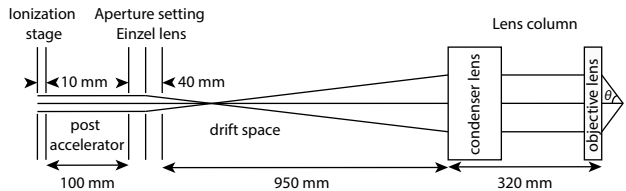}
\caption{Schematic view of the complete ion beam line incorporated in the GPT simulations for the probe size optimization.\label{beam_line}}
\end{figure}

The whole ion beam line is schematically shown in figure \ref{beam_line}. The ions are created in the beginning of the acceleration stage with the same initial distributions as described in section \ref{heating_simulations}. In the acceleration stage a variable electric field is applied to accelerate the ions. The ions are further accelerated to 30 keV in the post acceleration stage. They then enter the aperture setting einzel lens from which the outer plates are grounded and the inner plate is set at a variable positive voltage (0-30 kV) to reach the desired aperture angle. In the simulation the einzel lens is constructed from three infinitely thin conducting plates with a circular hole with a radius of 1 mm. The accelerator plates in the acceleration and post acceleration stage are constructed of such plates as well. After a drift space the ions reach the lens column. The electric fields of a realistic lens column were implemented in the simulations. This column basically consists of two einzel lenses; the condenser lens, which aims to collimate the beam and the objective lens, which is used to focus the beam to a small spot. All in all the simulation has four input variables: the acceleration field, the einzel lens voltage, the condenser lens voltage and the objective lens voltage. These parameters are varied by the genetic algorithm to find the pareto front of the problem.

\subsection{Results}\

The result of the genetic algorithm optimization is shown in figure \ref{d50vsI}a. For low currents, the simulated spot sizes are very similar to the model, including the current at which the transition from a spherical aberration dominated spot size to a chromatic aberration dominated spot size takes place. However, at larger currents the pareto front of the optimization deviates from the model, i.e. the simulations lead to smaller spot sizes than the analytical model predicts. As explained at the end of section \ref{calculations}, disorder induced heating is suppressed within a factor of two in the analytical model. However, at large currents the electric field that is needed to do so, see equation \ref{E(I)}, becomes so large that it is more beneficial to accept some disorder induced heating so that a lower electric field can be applied and the chromatic aberration becomes smaller.

This explanation is verified by figure \ref{d50vsI}b, which shows the electric field of the pareto solution as a function of the current in comparison with the electric field in the analytical model. At low currents the electric field is not of much importance, because the beam is limited by spherical aberrations. Therefore the electric field in the simulation deviates from the one in the model in this region. However, at a certain current the electric field rises and has approximately the same value as given by equation \ref{E(I)}. As can be seen this transition is not fluent in the simulation data. This is an artefact of the genetic optimization however. Since the problem has a relatively high number of input parameters it takes very long for the pareto solution to evolve to the actual optimum. The optimization shown here was stopped manually after a week.

From a certain current the electric field starts to deviate from equation \ref{E(I)}. This is the current for which not all disorder-induced heating is suppressed anymore, since this strategy leads to smaller spot sizes as shown in figure \ref{d50vsI}a. Since the genetic optimization was stopped manually without any well defined stopping criteria, the solution shown in figure \ref{d50vsI} possibly deviates from the actual optimal solution, i.e., there might be smaller spot sizes possible. 

\section{Conclusion}

The analytical calculations and particle tracing simulations of a complete and realistic ion beam line shown here predict that a FIB based on photo-ionizing a laser-intensified thermal atomic beam will outperform a LMIS based FIB in terms of spot size. By varying the electric field in which the ions are created for each beam current one can tune the trade-off between brightness and energy spread in order to get the optimum FIB performance. Nanometer size spots can be attained with a 30 keV beam of rubidium ions up to currents of a few pA. Due to the higher mass of rubidium atoms, better milling performance is expected than GFIS based FIBs, although this should be investigated in the future. At low currents the beam will be limited by a combination of spherical aberration and the beams brightness, while at higher currents chromatic aberration will be dominant. On the basis of the analytical result it can be concluded that the sensitivity of the final spot size to the initial flux density and temperature is low. Therefore less efficient laser cooling and compression than expected will not have dramatic consequences.

\begin{acknowledgments}
This research is supported by FEI Company, Pulsar Physics, Coherent Inc. and the Dutch Technology Foundation STW, applied science division of the ``Nederlandse Organisatie voor Wetenschappelijk Onderzoek (NWO)'' and the Technology Program of the Ministry of Economic Affairs.
\end{acknowledgments}



%
%

%


\bibliography{literature}

\end{document}